\documentclass[twocolumn,aps,showpacs,prb,tightenlines,amsmath,amssymb]{revtex4}
\usepackage{graphicx}
\usepackage{amssymb}
\usepackage{dcolumn} 
\usepackage{amsmath}
\usepackage{bm}
\usepackage{colordvi}
\usepackage{mathrsfs}
\newcommand{\bgreek}[1]{\mbox{\boldmath$#1$\unboldmath}}
\makeatletter

\newcommand{\Rmnum}[1]{\expandafter\@slowromancap\romannumeral #1@}
\makeatother

\begin{document} 

\title{Majorana fermions in semiconductor nanostructures with two wires connected through ring}
\author{B. Y. Sun} 
\author{M. W. Wu}
\thanks{Author to whom correspondence should be addressed}
\email{mwwu@ustc.edu.cn}
\affiliation{Hefei National Laboratory for Physical Sciences at
  Microscale and Department of Physics,
University of Science and Technology of China, Hefei,
  Anhui, 230026, China}
\date{\today}

\begin{abstract}
We investigate the Majorana fermions in a semiconductor nanostructure with two
wires connected through a ring. The nanostructure is mirror symmetric and in the proximity of a
superconductor. The Rashba spin-orbit coupling and a magnetic field parallel to the wires or
perpendicular to the ring are included. Moreover, a magnetic flux is applied through the center of
the ring, which makes the phase difference of the superconducting order parameters in the two wires
being zero or $\pi$ due to the fluxoid quantization and the thermodynamic
equilibrium of the supercurrent in the superconducting ring. If the phase
difference is $\pi$, two Majorana modes are shown to appear around the ring without
interacting with each other. On contrast, if the phase difference is zero, these
Majorana modes disappear and the states localized around the ring have finite 
energies. These states can be detected  via the conductance measurement
by connecting two normal leads to the wires and a third one directly to the
ring.  It is shown in the bias dependence of the differential conductance  from
one of the leads connected to the wire to the one connected directly to the ring
that the tunnelings through the Majorana modes (i.e., in the case with $\pi$ phase
difference) leads to two peaks very close to the zero bias, while the
tunneling through the states with finite energies (i.e., in the case with zero
phase difference) leads to peaks far away from the zero bias if the ring radius
is small. This difference for the cases with and without the Majorana modes
in small ring radius is distinct and hence can be used to identify the
Majorana modes. In addition, we also find that, because of the mirror inversion
symmetry of the nanostructure, the Andreev reflection through the lead
connected at the ring (which is along the inversion axis) is forbidden
around the zero bias if the magnetic flux is zero and the magnetic field is
parallel to the wires.
  
\end{abstract}


\maketitle
\section{INTRODUCTION}
In recent years, Majorana fermions have attracted immense attention due to
their possible applications in quantum
computation.\cite{Kitaev03,Sau11,Halperin12,Ivanov01,Read00,Sarma08,Alicea11}  
They are proposed to exist in various
systems\cite{Tanaka09,Tanaka12,Nakosai12,Linder10,Nakosai13,Kitaev03,Fu08,Ghaemi12,Sau10,Oreg10,Nadj-Perge13,Klinovaja13,Sau13,Stanescu13,Benjamin10}
and many experimental attempts have been devoted to identify
them.\cite{Sasaki11,Das12,Mourik12,Deng12,Rokhinson12,Churchill13,Williams12,Lee14} Among these
attempts, the zero-bias conductance
peak\cite{Bolech07,Law09,Yamakage12,Flensberg10,Prada12,Lin12,Diez12} is a characteristic
property from 
the Majorana modes which is often detected. This conductance measurement is easy
to perform and the zero-bias conductance peak has 
already been observed.\cite{Sasaki11,Das12,Mourik12,Deng12,Churchill13,Lee14} 
Nevertheless, these experimental results can not serve as decisive
evidence since the zero-bias conductance peak can appear due to many other
mechanisms.\cite{Chang13,Rainis13,Liu12,Kells12} Hence, to further identify the Majorana
fermions, the comprehension and detection of more transport properties at different Majorana
configurations are needed.

\begin{figure}[htb]
  \includegraphics[width=8.7cm]{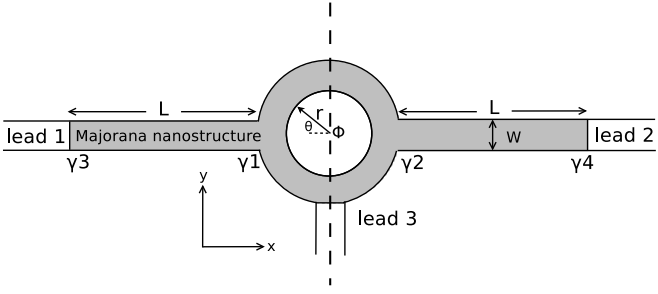}
  \caption{Schematic view of the Majorana nanostructure with two wires
    connected through a ring. The nanostructure is connected with three normal
      leads. Moreover, without the magnetic flux, 
      it has the mirror inversion symmetry with the inversion axis shown
      as the dashed line.}
  \label{figsw1}
\end{figure}

Among the works investigating Majorana nanostructures, the system of two connected nanostructures with different
order parameter phases is often investigated. It is found in various
materials\cite{Fu09,Lutchyn10,Kwon04,Xu14} that when this phase 
difference is $\pi$, two Majorana modes can appear around the
contacting point. Although they are close, these Majorana 
modes do not interact with each other, which is quit different from the result
that the Majorana fermions at the adjacent ends of the nanowire interact with each other
due to their wave-function overlapping.\cite{Sarma12} 
Inspired by these results, we propose a semiconductor nanostructure with two wires
connected through a ring as illustrated in Fig.~\ref{figsw1}. The nanostructure is in the proximity of a
superconductor. With a magnetic flux $\phi$ through the superconducting ring, the
variation of the supercurrent in the superconductor ring is governed by the 
fluxoid quantization and the thermodynamic equilibrium.\cite{Pientka13} Then, due to the
proximity effect, the order parameter in the semiconductor ring varies as\cite{Pientka13}
\begin{equation}
\Delta(\theta)=|\Delta_0|\exp(i[2\phi/\phi_0]\theta). \label{DeltaTheta}
\end{equation}
Here, $[x]$ denotes the integer closest to $x$ and $\phi_0=h/e$ is the flux quantum. Since the phase
difference of the order parameters between the left ($\theta=0$) and right ($\theta=\pi$) wires
is $[2\phi/\phi_0]\pi$, Majorana fermions are expected to appear around the ring when $[2\phi/\phi_0]$ is an odd
number. With these Majorana fermions, the differential
conductance through the lead at the ring (i.e., lead~3 in
Fig.~\ref{figsw1}) is expected to be different from those without these Majorana
fermions. Moreover, since $[2\phi/\phi_0]$ is an odd integer in a large range of
$\phi$ (e.g., $0.25\le\phi/\phi_0<0.75$), these Majorana modes should be robust
against the small variation of $\phi$.

In this work, we investigate the low energy states and the transport
properties of the semiconductor structure addressed above
(Fig.~\ref{figsw1}). The Rashba spin-orbit coupling and the Zeeman splitting
from a magnetic field along either ${\bf x}$ or ${\bf z}$-axis are included. A magnetic flux is
applied through the ring  which varies the phases of
the order parameters in the ring and the two wires [see Eq.~(\ref{DeltaTheta})]. We find that when the 
difference of phases in the wires is $\pi$, two Majorana modes
 indeed appear around the
ring. However, when the phase difference is zero, the states around
the ring have finite energies. When the ring radius is small, the 
properties of the differential conductance from leads 1
to 3 in these two cases are different. For the case with the Majorana  
modes, the interference of the Andreev
  reflections through these two Majorana modes leads to two
  differential conductance peaks very close to the zero bias in
 the bias dependence.
For the case without the Majorana modes, the localized
states around the ring  lead to resonant peaks far away
  from the zero bias.
 The different behaviors in the two cases under small
ring radius are easy to be distinguished from each other.
Moreover, since the phase difference
changes discontinuously from zero to $\pi$, the differential conductance peaks
also change {\em discontinuously} from the case without the Majorana modes to
those with them. This is a unique property which can be used to identify the
existence of the Majorana modes from the other mechanisms leading to the zero-bias
peak. Nevertheless, this detection is necessary to be carried out under
  small ring radius. This is because that the energies of the states
  around the ring under zero phase difference can be close to zero at a large
  ring radius, which makes the peaks difficult to distinguish from each other. Furthermore, we also
find that if the magnetic filed is along the {\bf x}-axis and the magnetic flux
is not applied, the Andreev reflection through lead 3 is zero around the zero
bias due to the mirror inversion symmetry.

This paper is organized as follows. In Sec.~{\Rmnum 2}, we set up the model and
lay out the formalism. In Sec.~{\Rmnum 3} the results obtained numerically  are
presented. We summarize in Sec.~{\Rmnum 4}.

\section{MODEL AND FORMALISM}
We investigate the semiconductor nanostructure shown in
Fig.~\ref{figsw1}. The Rashba spin-orbit coupling, proximity-induced
superconducting pairing and the Zeeman splitting from a magnetic field are
included and the width of the wire $w$ is assumed to be very small so that there
is only a single 1D mode occupied. Without the leads, the Hamiltonian is expressed as
\begin{equation}
\hat{H}_{\rm eff}=\hat{H}_{\rm rib}+\hat{H}_{\rm ring}+\hat{H}_{\rm SC}+\hat{H}_{\rm hop}=\hat{H}_0+\hat{H}_{\rm hop}.
\end{equation}
By discretizing the continuous Hamiltonian over a discrete lattice, the Hamiltonian under the
tight-binding approximation is obtained. For the Hamiltonian of the nanowire,\cite{YZhou14}
\begin{eqnarray}
&&\hspace{-0.8cm}  \hat{H}_{\rm rib}=\sum_{\sigma\sigma'\mu}\sum^L_{i=1}a_0[{\sigma}^n_{\sigma,\sigma'} 
  V_n+(V_{i}- \mu)\delta_{\sigma,\sigma'}] c_{i\sigma\mu}^\dagger c_{i\sigma'\mu} \nonumber \\
&& \hspace{-0.9cm}\mbox{}  -\hspace{-0.1cm} \sum_{\langle i,j \rangle \sigma\mu}a_0 t c_{i\sigma\mu}^\dagger c_{j\sigma\mu} 
  +i E_R \sum_{\langle i,j \rangle\sigma\sigma'\mu} a_0
  v^x_{ij}\sigma^y_{\sigma \sigma'} c_{i \sigma\mu}^\dagger c_{j \sigma'\mu}.
\end{eqnarray}
Here, $V_n$ (with $n=x$ or $z$) stands for the Zeeman splitting from the magnetic field along the ${\bf
  n}$-axis; ${\bgreek \sigma}$ stand for the Pauli
matrices; $L$ is the number of sites at each wire; $\mu=-1$ $(1)$ stands for the left
(right) wire; $t=\hbar/(2m^\ast a_0^2)$  represents the hopping energy with $a_0$ denoting the 
distance between the nearest neighbors in the wire; $V_{i}=2t$ stands for the on-site energy;
$\langle i,j \rangle$ represents a pair of the nearest neighbors;
$v_{ij}^x={\bf e}_x \cdot {\bf d}_{ij}$ with 
${\bf d}_{ij}=({\bf r}_i-{\bf r}_j)/|{\bf r}_i-{\bf r}_j|$;  
$E_R$ is the Rashba spin-orbit-coupling constant.
The Hamiltonian of the ring reads\cite{Meijer02,Joibari13,Pientka13}
\begin{eqnarray}
  \hat{H}_{\rm ring}\hspace{-0.2cm}&=&\hspace{-0.2cm}\sum_{\sigma\sigma^\prime}\sum^{M-1}_{i=0}\frac{2\pi r}{M}[ {\sigma}^n_{\sigma,\sigma'} 
  V_{n}+(V^r_{i}- \mu)\delta_{\sigma,\sigma'}] c_{i\sigma,0}^\dagger c_{i\sigma',0} \nonumber \\
&& \hspace{-1.2cm}\mbox{}- \sum_{\langle i,j \rangle \sigma}\frac{2\pi r}{M} t_r c_{i\sigma,0}^\dagger c_{j\sigma,0}+{i E^r_R} \sum_{\langle i,j \rangle\sigma\sigma'} \frac{2\pi r}{M}
  v^\theta_{ij}c_{i \sigma,0}^\dagger c_{j \sigma',0}\nonumber \\ && \hspace{-1.2cm}\mbox{}\times [(\sin\theta_i+\sin\theta_j)\sigma^y_{\sigma
    \sigma'}
+(\cos\theta_i+\cos\theta_j)\sigma^x_{\sigma\sigma'}],
\end{eqnarray}
in which $M$ is the number of sites in the ring; $c_{i\sigma,0}$ is the
annihilation operator for the electron with spin $\sigma$ at the site $i$ of
the ring; $V^r_i=2t(M a_0)^2/(2\pi r)^2$; $t_r=t(M a_0)^2/(2\pi r)^2e^{i2\pi\phi v^\theta_{ij}/({N\phi_0})}$;
$E^r_R=E_Ra_0Me^{i2\pi\phi v^\theta_{ij}/({N\phi_0})}/(4r\pi)$ and 
$v^\theta_{ij}=\sin[2\pi(j-i)/M]/|\sin[2\pi(j-i)/M]|$. 
The Hamiltonian for the hopping between the ring and the wires is
\begin{eqnarray}
  \hat{H}_{\rm hop}\hspace{-0.2cm}&=&\hspace{-0.1cm}-\sum_{\sigma}a_0 t c_{L,\sigma,-1}^\dagger c_{0\sigma,0} 
+{i E_R} \sum_{\sigma\sigma'}a_0 \sigma^y_{\sigma \sigma'} c_{L,\sigma,-1}^\dagger
c_{0\sigma',0}\nonumber \\
&&\hspace{-1.3cm}\mbox{}-\sum_{\sigma}a_0 t c_{\frac{M}{2},\sigma,0}^\dagger c_{0\sigma,1} 
+{i E_R} \sum_{\sigma\sigma'}a_0 \sigma^y_{\sigma \sigma'} c_{\frac{M}{2},\sigma,0}^\dagger
c_{0\sigma',1}+{\rm H.c.}.\nonumber\\
\end{eqnarray}
As for the  pairing potential, it is given by 
\begin{equation}
  \hat{H}_{\rm SC}=\sum_{n\mu}a_0 \Delta_{n,\mu} c_{n\uparrow\mu}^\dagger
  c_{n\downarrow\mu}^\dagger+\sum^{M-1}_{n=0}\frac{2\pi r}{M} \Delta_{n,0} c_{n\uparrow0}^\dagger
  c_{n\downarrow0}^\dagger + {\rm H. c.},
\end{equation}
with $\Delta_{n,-1}=|\Delta_0|$, $\Delta_{n,1}=|\Delta_0|\exp(i[2\phi/\phi_0]\pi)$
and $\Delta_{n,0}=|\Delta_0|\exp(i[2\phi/\phi_0]2\pi n/M)$.

Using the convention of the Nambu spinors,
$\Psi^\dag_{i\mu}=(c^\dag_{i\uparrow\mu},c^\dag_{i\downarrow\mu},c_{i\downarrow\mu},-c_{i\uparrow\mu})$,
this Hamiltonian is rewritten into
\begin{equation} 
\hat{H}_{\rm eff}=\frac{1}{2}\sum_{i\mu j\nu}\Psi^\dag_{i\mu}H_{\rm BdG}(i\mu,j\nu)\Psi_{j\nu},
\end{equation}
with 
\begin{equation}
  {H}_{\rm BdG}(i\mu,j\nu)=\begin{pmatrix} \hat{H}_{0}(i\mu,j\nu) && \Delta_{i,\mu} \delta_{ij}\delta_{\mu,\nu} \\
  \Delta^*_{i,\mu} \delta_{ij}\delta_{\mu\nu} && -{\sigma}_y \hat{H}_{0}^\ast(i\mu,j\nu) {\sigma}_y
  \end{pmatrix}
\label{BdG}
\end{equation}
being the Bogoliubov-de Gennes (BdG) Hamiltonian.\cite{BdG_original}
From this $H_{\rm BdG}$, the eigenstate
$\psi^n_{i\mu}=(u^n_{i\uparrow\mu},u^n_{i\downarrow\mu},v^n_{i\downarrow\mu},v^n_{i\uparrow\mu})^T$
for the $n$-th state can be obtained numerically  
under the normalization condition of the wave function $\sum_{i\sigma,\pm1}a_0
(|u^n_{i\sigma,\pm1}|^2+|v^n_{i\sigma,\pm1}|^2)+\sum_{i\sigma}(2\pi r/M)(|u^n_{i\sigma,0}|^2+|v^n_{i\sigma,0}|^2)=1$.

To calculate the transport property of this nanostructure, we connect
normal leads to it as shown in Fig.~\ref{figsw1}. The Hamiltonian of the
lead is given by 
\begin{eqnarray}
H_\eta&=&\sum_{i\sigma\sigma^\prime}a_0[{\sigma}^n_{\sigma,\sigma'} V_{n}+(V_{i}-
\mu-\mu_\eta)\delta_{\sigma,\sigma'}] d_{i\sigma}^{\eta\dagger}
d^{\eta}_{i\sigma'}\nonumber \\ &&\mbox{}-\sum_{\langle i,j \rangle \sigma} a_0t d_{i\sigma}^{\eta\dagger} d^{\eta}_{j\sigma},
\end{eqnarray}
in which $\eta=1$, 2,
3 represent the leads shown in Fig.~\ref{figsw1} and $\mu_\eta$ stands for the difference of the chemical potential between
lead $\eta$ and the nanostructure. The hopping between the leads and the nanostructure is 
\begin{equation}
H_T=\sum_{\eta\sigma\sigma'}a_0T^\eta_{i\sigma,j\sigma'}d^{\eta\dag}_{\eta
  i\sigma}c_{j\sigma'}+{\rm H.c.},
\end{equation}
with $T^\eta_{i\sigma,j\sigma'}=-t\delta_{\sigma\sigma'}$, where $i$ and
$j$ stand for the contacting points between the leads and the nanostructure.

We investigate the conductance of the nanostructure at zero temperature. Then, the
current through lead $\eta$ is given by\cite{YZhou14}
\begin{equation}
  I_\eta =\frac{e}{h} \sum_{\eta'\beta}
  \int_{\chi_\beta\mu_{\eta'}}^{\mu_\eta} d\varepsilon \;
  P_{\eta\eta'}^{e\beta}(\varepsilon),
\label{current}
\end{equation}
with $\chi_\beta=1$ $(-1)$ for $\beta=e$ ($h$) and 
\begin{equation}
  P^{\alpha\beta}_{\eta\eta'}(\varepsilon)={\rm Tr} \left\{
    \hat{G}^{r}(\varepsilon) \hat{\Gamma}^{\beta}_{\eta'}(\varepsilon) 
    \hat{G}^{a}(\varepsilon)
    \hat{\Gamma}^{\alpha}_{\eta}(\varepsilon) \right\}. 
\label{Trans}
\end{equation}
Here, $\hat{G}^{r,a}(\varepsilon)$ are the retarded and advanced Green's functions 
in the nanostructure connected with the leads; $\hat{\Gamma}^{\alpha}_{\eta}(\varepsilon)$ is the
self-energy from the electric ($\alpha=e$) or hole part ($\alpha=h$) of the lead
$\eta$.\cite{YZhou14}

With the bias applied to the nanostructure, the chemical potentials of the leads
are shifted according to $V_{12}=\mu_1-\mu_2$ and $V_{13}=\mu_1-\mu_3$. Then,
with the further constraint of current conservation 
\begin{equation}
\sum_\eta I_\eta=0,
\label{CurrentConservation}
\end{equation}
the differential conductance is obtained. In this work, we investigate the
differential conductance between two leads without current flowing through the
remaining one, i.e., $G_{12}=\frac{dI_1}{dV_{12}}\big|_{I_3=0}$ and
$G_{13}=\frac{dI_1}{dV_{13}}\big|_{I_2=0}$. It is noted that when the wires are very
long, the transmission between different leads becomes negligible around the
zero bias due to the superconducting gap. In this case, only the Andreev
reflection contributes and whence Eq.~(\ref{current}) becomes 
\begin{equation}
  I_\eta =\frac{e}{h} \int_{-\mu_{\eta}}^{\mu_\eta} d\varepsilon \;
  P_{\eta\eta}^{eh}(\varepsilon).
\label{currentAnd}
\end{equation}
Then, the differential conductance is given by 
\begin{equation}
G_{1\eta}=-e[P_{\eta\eta}^{eh}(\mu_\eta)+P_{\eta\eta}^{eh}(-\mu_\eta)]{\partial_{V_{1\eta}}\mu_\eta}/h.
\label{G13}
\end{equation}
Here, $\mu_\eta$ and ${\partial_{V_{13}}\mu_\eta}$ are determined by the current
conservation and $V_{1\eta}=\mu_1-\mu_\eta$.

\section{RESULTS}
We investigate the low energy states and the transport properties of the
nanostructure. In our computation,
$\Delta_0=E_R=0.2t$, $L=200$, $M=256$ and $V_n=0.8t$ unless otherwise
specified. It is noted that, with these parameters, the 
nanostructure is in the topological nontrivial regime, i.e.,
$\sqrt{\Delta^2+\mu^2}<|V_n|<\sqrt{\Delta^2+(\mu-4t)^2}$.\cite{YZhou14} Hence, the Majorana
modes are expected to appear at the ends of the nanostructure.

\subsection{Majorana states and energy spectrum}
\begin{figure}[htb]
  \includegraphics[width=9.1cm]{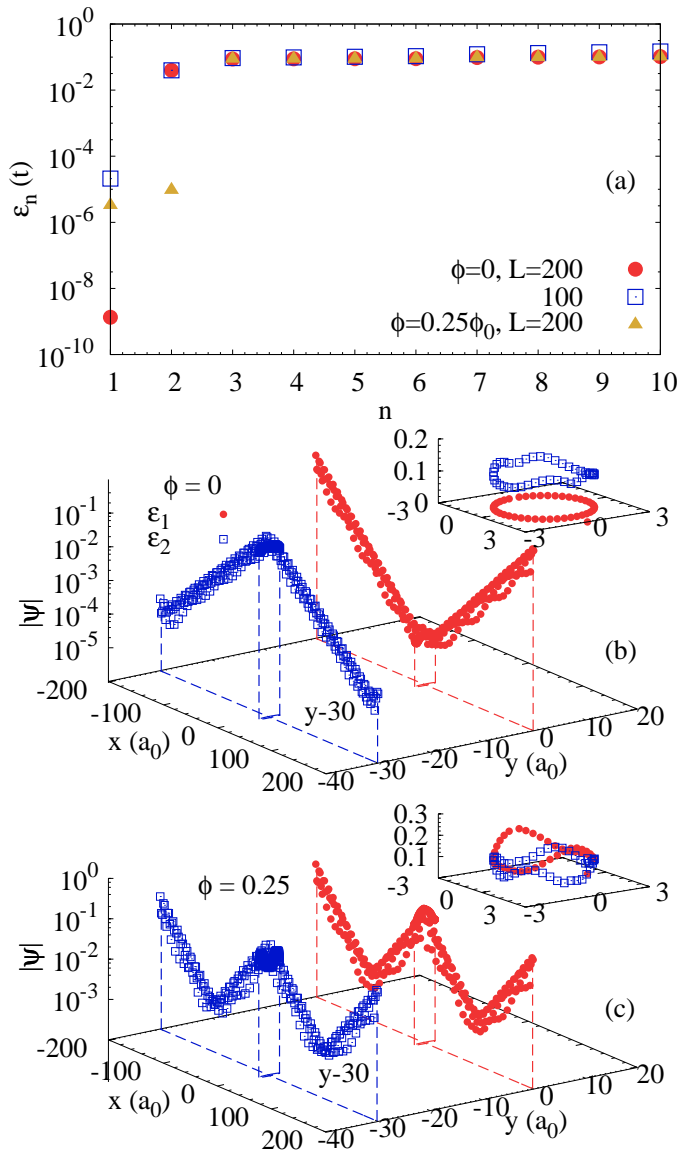}
  \caption{(Color online) (a) Low energy spectra with different wire lengths
    for $\phi=0$ and $0.25\phi_0$, respectively. $n$ labels the 
    eigenvalues of $H_{\rm BdG}$ starting with zero energy.
    (b) and (c) Magnitudes of the wave functions of the lowest two
    states ($\epsilon_1$ and $\epsilon_2$) for $\phi=0$ and $0.25\phi_0$,
    respectively. Those of the second lowest states (blue squares) are shifted by $30a_0$
    along the ${\bf y}$-axis for clarity. The insets zoom the wave functions at
    the ring of the nanostructure. The magnetic field is along the ${\bf
      z}$-axis and the radius of the ring is $2a_0$.}  
  \label{figsw2}
\end{figure}
In this section, we present the numerical results of the low-energy spectrum and
the eigenstates. The radius of the ring is taken to be $r=2a_0$. Due to the
particle-hole symmetry of $H_{\rm BdG}$, we only investigate the results with
positive eigenvalues. The low-energy spectra of the nanostructure 
under different magnetic fluxes are plotted in Fig.~\ref{figsw2}(a) and the
magnitudes of the wave functions for the lowest two states are shown in
Figs.~\ref{figsw2}(b) and (c). It is noted that the energy spectra
 with the Zeeman splitting
from a magnetic field along either ${\bf x}$- or ${\bf z}$-axis are
similar. Hence, we only plot the results with a magnetic
field in ${\bf z}$-axis in Fig.~\ref{figsw2}.

From Fig.~\ref{figsw2}(a), one finds that for 
the case with the magnetic flux $\phi=0$ (red dots), the eigenvalue
$\varepsilon_1$ is extremely small while the other eigenvalues are much
higher. Moreover, from the distribution of the wave functions shown in
Fig.~\ref{figsw2}(b), it is found that the lowest state mainly distributes at
the ends of the nanostructure and the second lowest state mainly distributes around the
ring. These behaviors can be easily understood as follows. If the ring is removed
from the nanostructure, it is well known that Majorana modes exist at the ends of the
left and right wires (marked as $\gamma_1$ to $\gamma_4$ in Fig.~\ref{figsw1}). With a 
ring connecting these two wires, the Majorana modes 
close to the ring in each wire (i.e., $\gamma_1$ and $\gamma_2$ in
Fig.~\ref{figsw1}) interact with each other and form the second lowest
state. For the remaining Majorana modes ($\gamma_3$ and $\gamma_4$ in
Fig.~\ref{figsw1}), they compose the lowest eigenstate with the corresponding
non-zero eigenvalue coming from the interactions between them. To further
confirm this, we also calculate the case with a shorter wire length $L=100$
(blue squares). It is shown that $\varepsilon_1$ increases markedly while
$\varepsilon_2$ remains almost unchanged, which is consistent with the feature
of the interacting Majorana fermions.

We further investigate the influence of the magnetic flux by increasing $\phi$. The results are
similar to the case with $\phi=0$ until $\phi=0.25\phi_0$ where $[2\phi/\phi_0]=1$ and
$\Delta_{n,-1}=-\Delta_{n,1}$. The low energy spectrum is plotted as yellow
triangles in Fig.~\ref{figsw2}(a). It is shown that both the lowest and the second
lowest energies are very close to zero. Furthermore, from the magnitudes of the
wave functions for the lowest two eigenstates shown in 
Fig.~\ref{figsw2}(c), it is found that the two wave functions distribute both around the
ring and at the ends of the nanostructure. These behaviors indicate the
existence of the Majorana modes around the ring. These Majorana modes interact
with those at the ends of the nanostructure ($\gamma_3$ and $\gamma_4$) which leads to
the small nonzero eigenvalues (i.e., $\varepsilon_1$ and
$\varepsilon_2$). Since the interaction between them decreases 
exponentially with the increase of their distance, we also calculate the
case with $L=300$, which leads to much lower $\varepsilon_1$ ($2.5\times10^{-8}t$) and
$\varepsilon_2$ ($7.7\times10^{-8}t$). This behavior confirms the existence of
the Majorana fermions around the ring. It is noted that this result is similar to those
appearing in two connected nanostructures with a $\pi$ phase difference of the
order parameters investigated in the literature.\cite{Fu09,Lutchyn10,Kwon04,Xu14}

\subsection{Electric conductance}
\begin{figure}[htb]
  \includegraphics[width=8.5cm]{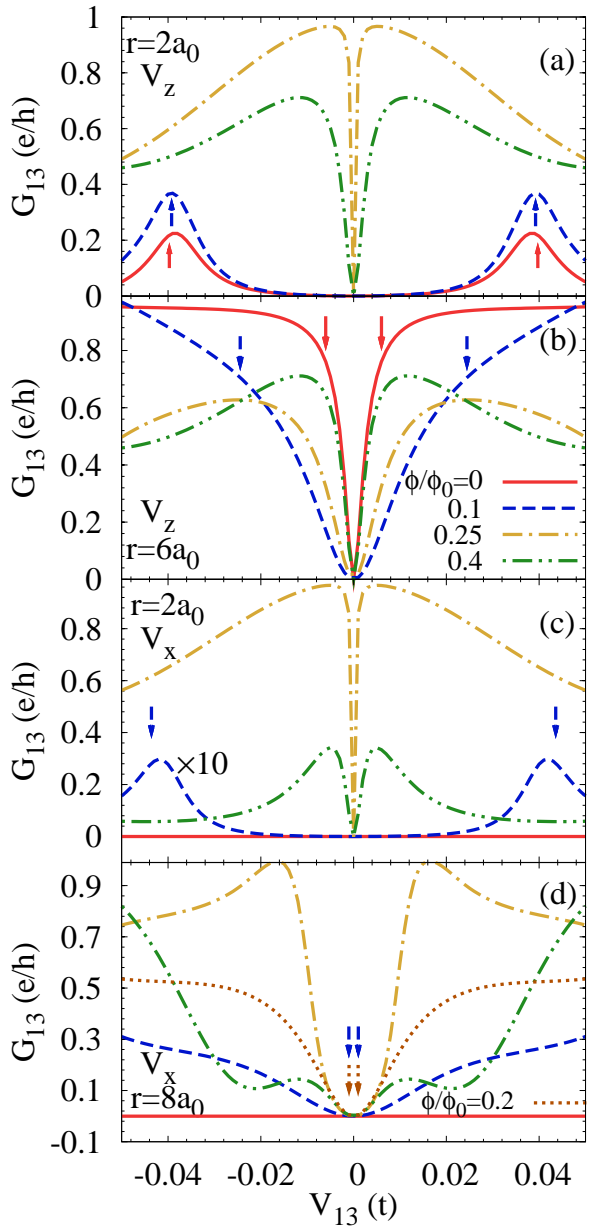}
  \caption{(Color online) Differential conductance for cases with the magnetic
    field along the ${\bf z}$-axis with (a) $r=2a_0$ and (b) $r=6a_0$ as well as the case
    along the ${\bf x}$-axis with (c) $r=2a_0$ and (d) $r=8a_0$ under
    different magnetic fluxes. The results with $\phi=0.1\phi_0$ (blue dashed curve) in (c) is
    enlarged by 10 for clarity. The arrows indicate the energies of the corresponding second
    lowest states of the nanostructure.}
  \label{figsw3}
\end{figure}
In this section, we investigate the influence of the Majorana modes on
the conductance by varying the magnetic flux $\phi$. Here, the wires are long enough and 
only the Andreev reflection contributes to the transport. Thus, the
  differential conductance through each lead is mainly determined by the states around the lead
  which helps to show their influence clearer. It is noted that
the property of the differential conductance between leads 1 and 2, i.e., $G_{12}$, is
mainly determined by the Majorana modes at the ends of the nanostructure ($\gamma_3$ and
$\gamma_4$ shown in Fig.~\ref{figsw1}), which is
similar to previous works in the literature.\cite{YZhou14,Lim12} Hence, it is not shown in the
figure. In this work we only concentrate on the differential conductance between
leads 1 and 3, i.e., $G_{13}$.

We first focus on the case with a short ring radius (i.e., $r=2a_0$). 
For the case with $\phi=0.25\phi_0$ where the phase difference of the order
parameters in the two wires is $\pi$ (i.e., with the Majorana modes around the
ring), two conductance peaks appear in the bias voltage $V_{13}$ dependence
which are very close to the zero bias (i.e., $V_{13}=0$) 
and hence lead to a sharp valley between them as shown in Fig.~\ref{figsw3}(a). With the
increase of $\phi$ to $0.4\phi_0$, where the phase difference is still $\pi$,  the
valley is shown to become milder. Nevertheless, the corresponding peaks are still close to the
zero bias. On contrast, for the case with zero phase difference (i.e.,
without any Majorana modes around the ring), e.g., $\phi=0$ and $0.1\phi_0$, the
peaks appearing in the bias dependence become close to the energies of the
second lowest states (marked as arrows) which are far away from zero.
This behavior is quite different from that with the Majorana modes and hence can
be used to distinguish the two cases. Moreover, since the 
phase difference varies discontinuously from zero to $\pi$, the 
conductance peaks without the Majorana modes also change to the ones
with them {\em discontinuously}. This is a unique property
of the Majorana modes which helps to exclude the other mechanisms leading to
the zero-bias peaks.

Here, we give an analytical investigation on the properties of the differential conductance
peaks in the cases with and without the Majorana modes around the ring. The
differential conductance $G_{13}$ is determined by the Andreev reflections through
leads 1 and 3 under the constraint of current conservation (which determines
$\mu_1$ and $\mu_3$ incorporating with $V_{13}=\mu_1-\mu_3$). For the Andreev
reflection through lead 1, it is strong due to the Majorana mode near lead
1 (i.e., $\gamma_3$) and the corresponding $P^{eh}_{11}$ is almost 1 in our parameter range. On the
contrary, the Andreev reflection through lead 3 varies largely and the
corresponding $P^{eh}_{33}$ is found to be zero around the zero bias. Therefore,
due to the current conversation [Eqs.~(\ref{CurrentConservation}) and
(\ref{currentAnd})] and $V_{13}=\mu_1-\mu_3$, $\mu_3$ is close to 
$-V_{13}$ ($\partial_{V_{1\eta}}\mu_\eta$ is close to $-1$) and the differential
conductance is mainly determined by the Andreev reflection through lead 3 [i.e.,
$P_{\eta\eta}^{eh}(\mu_3)+P_{\eta\eta}^{eh}(-\mu_3)$, see
Eq.~(\ref{G13})]. With this understanding, the conductance peaks close to the
zero bias in the case with $\pi$ phase difference are understood to 
come from the Andreev reflection through the Majorana modes around the ring. And
the peaks close to the energies of the second lowest states in the case with zero phase
difference are understood to come from the resonance Andreev reflection through
these subgap states [which are also localized around the ring, see Fig.~\ref{figsw2}
(b)]. It is noted that both the contributions of the Majorana modes (with $\pi$ phase
difference) and the second lowest states (with zero phase difference) can be
studied formally by investigating the Andreev reflection through states
$\psi^n$ and $\psi^{-n}$ with $\psi^{-n}={\psi^n}^\dagger$ due to the
partial-hole symmetry. Here, by setting $\psi^n$ being the fermion mode composed
by the Majorana modes $\gamma_{r1}$ and $\gamma_{r2}$ around the ring [i.e.,
$\psi^n=(\gamma_{r1}+i\gamma_{r2})/\sqrt{2}$] and setting the corresponding
energies $\varepsilon_{n}=\varepsilon_{-n}=0$, the result
describes the tunneling through the two Majorana modes. Meanwhile, by setting
$\psi^n$ being the second lowest state (i.e., $n=2$) and using
$\varepsilon_2=-\varepsilon_{-2}\ne0$, it describes the tunneling through the
second lowest states.

Since the lead 3 is connected to the nanostructure at one point, the couplings
between these two states $\psi^n$ and $\psi^{-n}$ to the lead 3 are determined by
their wave functions at the connecting point which are described by
$\psi^n=(u_\uparrow,u_\downarrow,v_\downarrow,v_\uparrow)^T$ and
$\psi^{-n}={\psi^{n}}^\dagger$ in the Nambu spinors. Then the approximate formula
is given as (see Appendix) 
\begin{equation}
P^{eh}(\varepsilon)={64|u_{\downarrow}|^2|v_{\downarrow}|^2\Gamma^e_L\Gamma^h_L\varepsilon^2}/{D(\varepsilon)}.
\label{PehZ}
\end{equation}
Here,
$D(\varepsilon)=[4(\varepsilon^2_n-\varepsilon^2)+\Gamma^e_L\Gamma^h_L(|u_{\downarrow}|^2-|v_{\downarrow}|^2)^2]^2+4[(\Gamma^e_L+\Gamma^h_L)\varepsilon(|u_{\downarrow}|^2+|v_{\downarrow}|^2)+(\Gamma^e_L-\Gamma^h_L)\varepsilon_n(|u_\downarrow|^2-|v_{\downarrow}|^2)]^2$. From this equation,
$P^{eh}(0)$ is found to be zero which explains the valley in our
results. Moreover, under the approximation $\Gamma^e_L=\Gamma^h_L$ (valid around
the zero bias), one further finds that the peaks locate at
$\pm\sqrt{\varepsilon^2_n+(|u_\uparrow|^2-|v_\uparrow|^2)^2\Gamma^{e}_L\Gamma^{h}_L/4}$. From 
this result, one understands that both $|u_\uparrow|^2-|v_\uparrow|^2$
and the energy of the state $\varepsilon_n$ influence the location of the corresponding peaks (as
well as the steepness of the valley between them). Nevertheless, if the energy $\varepsilon_n$ of the second 
lowest state in the case with zero phase difference (e.g., $\phi=0$ and $0.1\phi_0$) is much larger than
$\sqrt{(|u_\uparrow|^2-|v_\uparrow|^2)^2\Gamma^{e}_L\Gamma^{h}_L/4}$, the corresponding
peaks will be far away from those due to the Majorana modes (i.e.,
$\varepsilon_n=0$ in the case with $\pi$ phase difference, e.g.,
$\phi=0.25\phi_0$ and $0.4\phi_0$) as shown in Fig.~\ref{figsw3}(a).

\begin{figure}[htb]
  \includegraphics[width=8.5cm]{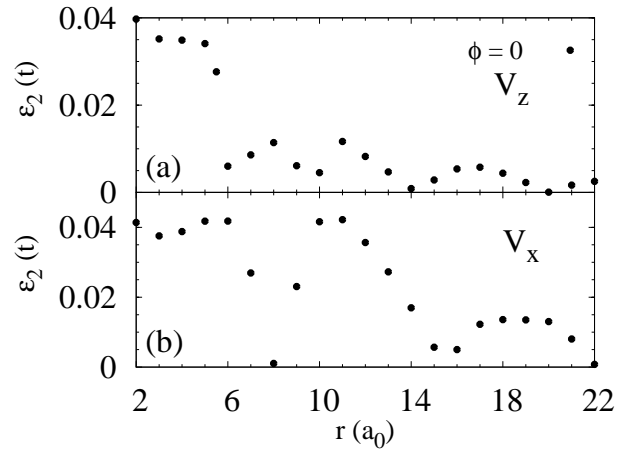}
  \caption{The ring radius dependence of the energy of the
    second lowest state $\varepsilon_2$ with the Zeeman splitting from the
    magnetic field along the (a) {\bf z}- and (b) {\bf x}-axes. Here, the
    magnetic flux $\phi=0$. }
  \label{figsw4}
\end{figure}

With this understanding, we further investigate the case with a different
ring radius. The variation of the ring radius changes the couplings between the
states in different wires. As a result, the energies of the second 
lowest states in the case with zero phase difference can be close to
zero, making the location of the peaks (or the steepness of the
corresponding valley) difficult to be distinguished from
those due to the Majorana modes. This can be seen in Fig.~\ref{figsw3}(b) in which the differential conductance 
$G_{13}$ is plotted against the bias voltage $V_{13}$ with $r=6a_0$. For the
case with zero phase difference (i.e., without any Majorana modes), it is shown
that the second lowest eigenvalues (marked by the 
arrows) can be very close to zero (e.g., $\phi=0$). Then, the steepness of
the valley is mainly determined by $|u_\uparrow|^2-|v_\uparrow|^2$ as the case
due to the Majorana modes. As a result, the conductance valleys with (e.g.,
$\phi=0.4\phi_0$) and without (e.g., $\phi=0$)\cite{Morestates} the Majorana modes can be
similar. This makes the two cases difficult to be distinguished from each
other. These results suggest the importance to rule out the second lowest
states in the case with zero phase difference. Thus, we further show the energies of the second lowest
states at $\phi=0$ as the function of the ring radius in
Fig.~\ref{figsw4}(a). It is shown that, with the increase of the ring radius,
the energy of the second lowest state oscillates and tends to
decrease. Nevertheless, the energies in the cases with $r<5a_0$ are always
large. This indicates that, to identify the existence of the Majorana modes
around the ring, it is necessary to detect in the case with small ring radius
($<5a_0$).

In addition to the case with a magnetic field along the ${\bf z}$-axis, we
further investigate the case with the magnetic field along the ${\bf 
  x}$-axis as shown in Figs.~\ref{figsw3}(c) and (d). Since the properties of the energy spectra for these two magnetic
field directions are similar (see Fig.~\ref{figsw4}), the corresponding
conductances are also expected to be similar as confirmed by the cases with the 
magnetic fluxes under
both small [$r=2a_0$ shown in Fig.~\ref{figsw3}(c)] and large [$r=8a_0$ shown in
Fig.~\ref{figsw3}(d)] ring radii. Nevertheless, if the magnetic flux is not
applied (i.e., $\phi=0$),  it is shown in Figs.~\ref{figsw3}(c) and (d) that the 
corresponding differential conductances keep zero in our parameter range. This
special property comes from the mirror inversion symmetry of the nanostructure
which is understood as follows.

With the magnetic field along the ${\bf x}$-axis applied to the nanostructure, the spin degeneracy in
the lead is split into $\sigma^+_x$ and $\sigma^-_x$. Since the Zeeman splitting is larger than the
chemical potential, only the lower eigenstate $\sigma^-_x$ in the lead contributes to the
conductance. This eigenstate only couples the state in the nanostructure with spin parallel to it,
whose component in the corresponding wave function $\psi^n$ is
$(u_{\uparrow}-u_{\downarrow})/\sqrt{2}$ for the electron part and 
$(v_{\uparrow}+v_{\downarrow})/\sqrt{2}$ for the hole part. Then, the Andreev
reflection is understood to be proportional to
$(u_{\uparrow}-u_{\downarrow})(v_{\uparrow}+v_{\downarrow})$ (see also Appendix). 
For our nanostructure without the magnetic flux, the Hamiltonian is 
invariant under the mirror inversion with the inversion axis along the lead 3 (the corresponding operator is 
${\bgreek\pi}\sigma_x$, with ${\bgreek \pi}^\dagger {\bf x}{\bgreek \pi}=-{\bf
  x}$). Then, at the point contacting with lead 3, each non-degenerate eigenstate of the nanostructure has
to satisfy the relation $\psi^n=\pm\sigma_x\psi^n$, i.e.,
$(u_{\uparrow},u_{\downarrow},v_{\downarrow},v_{\uparrow})^T=\pm(u_{\downarrow},u_{\uparrow},v_{\uparrow},v_{\downarrow})^T$. As a result, $u_{\uparrow}-u_{\downarrow}=0$ or $v_{\uparrow}+v_{\downarrow}=0$. Hence the
Andreev reflection through lead 3 is forbidden, leading to the 
zero $G_{13}$.

\section{SUMMARY}
In summary, we have investigated the Majorana fermions in a semiconductor nanostructure
with two long wires connected through a ring. The nanostructure is in the proximity of a superconductor and
the Rashba spin-orbit coupling, proximity-induced superconducting pairing and
the Zeeman splitting from a magnetic field are included. A magnetic flux is applied through the
center of the ring. Then, due to the fluxoid quantization and the thermodynamic
equilibrium of the supercurrent in the superconducting ring,\cite{Pientka13}
the phase difference between the order parameters of the two semiconductor wires
is zero or $\pi$. We show that when it is zero, the states in the two wires are
coupled through the ring and the eigenstates localized around the ring have
finite energies (i.e., there is no Majorana modes around the ring). Nevertheless, when the
phase difference becomes $\pi$, two Majorana states appear around the ring
without interacting with each other.

We further investigate the transport property of the nanostructure by
  connecting two normal leads to the wires and a third one directly to the
  ring. The low bias differential conductance between one of the leads connected
  at the wire and the one at the ring is mainly limited by the small Andreev
  reflection through the lead connected at the ring, due to the current
  conversation.  Hence, in both cases, i.e., with and without the Majorana modes
  around the ring,  this differential conductance can show very distinct 
features.
  In the case with the Majorana modes (i.e., with $\pi$ phase difference), the
  bias dependence of the differential conductance exhibits two peaks very close to the zero bias due
  to the interference of the Andreev reflections through these two Majorana modes. On contrast, 
  in the case without the Majorana modes (i.e., with zero phase
  difference), it shows peaks far away from the zero bias if the ring radius is
  small. This is due to the resonant Andreev reflections through the localized
  states with finite energies. The behaviors in these two cases under small ring
  radius can be easily distinguished from each other. This is a 
  unique property of the Majorana modes which helps to exclude the other
  mechanisms leading to the zero-bias peak. Nevertheless, we also point out that
if the ring radius becomes large, the energies 
of the states in the case with the
 zero phase difference can be very close to zero, which makes the corresponding
  peaks difficult to be distinguished from those due to the Majorana
  modes in the case with $\pi$ phase difference. Hence, to identify the Majorana
  modes, small ring radius is a must.

In addition, we also find that due to the mirror symmetry of the nanostructure, 
the Andreev reflection through the lead connected
at the ring (which is along the inversion axis) is forbidden around the zero
bias if the magnetic field is parallel to the wires and the magnetic flux is
absent.

\appendix*
\section{Interference effect of the tunneling through two Majorana modes}
In this section, we present the interference effect of the Andreev reflections
through two states $\psi^n$ and $\psi^{-n}$. Due to the partial-hole 
symmetry, $\psi^{n\dag}=\psi^{-n}$ and their corresponding energies satisfy
$\varepsilon_{n}=-\varepsilon_{-n}$. Then, the Green function is written as 
\begin{eqnarray}
  \hat{G}^{r}(\varepsilon)&=&\left[\varepsilon-\begin{pmatrix} \varepsilon_{n} & 
      0\\ 0 & -\varepsilon_{n}\end{pmatrix}
    +\frac{i}{2}(\Gamma_e+\Gamma_h) \right]^{-1},
  \label{G_rApp} 
\end{eqnarray}
in which $-i\Gamma_e/2$ stands for the self-energy from the electron
part of the lead and $-i\Gamma_h/2$ from the hole part. They are expressed as 
\begin{equation}
  -\frac{i}{2}\Gamma_{e/h}(\varepsilon) =TG^{e/h}_LT^\dag,
  \label{GammaApp}
\end{equation}
with $G^{e}_L$ and $G^{h}_L$ standing for the electron and hole Green functions in
the lead and $T$ representing the hopping matrix from the two states to the lead.

For the case with the magnetic field along the ${\bf z}$-axis, considering that
the Majorana modes appear only when the Zeeman
splitting is larger than $\sqrt{\mu^2+|\Delta_0|^2}$, the Green functions in the
lead can be expressed as 
\begin{equation}
G^{e}_L=\begin{pmatrix} 0&0&0&0\\0&-\frac{i}{2}\Gamma^e_L&0&0\\0&0&0&0\\0&0&0&0
\end{pmatrix},\ \mbox{and}\ G^{h}_L=\begin{pmatrix} 0&0&0&0\\0&0&0&0\\0&0&-\frac{i}{2}\Gamma^h_L&0\\0&0&0&0
\end{pmatrix}\nonumber
\end{equation}
in the Nambu spinor basis. On the other hand, for the case with the magnetic field
along the ${\bf x}$-axis, they are expressed as 
\begin{eqnarray}
G^{e}_L=-\frac{i}{4}\begin{pmatrix} \Gamma^e_L&-\Gamma^e_L&0&0\\-\Gamma^e_L&\Gamma^e_L&0&0\\0&0&0&0\\0&0&0&0
\end{pmatrix},\nonumber\\ \ \mbox{and}\ 
 G^{h}_L=-\frac{i}{4}\begin{pmatrix} 0&0&0&0\\0&0&0&0\\0&0&\Gamma^h_L&\Gamma^h_L\\0&0&\Gamma^h_L&\Gamma^h_L
\end{pmatrix}\nonumber.
\end{eqnarray}
As for the hopping matrix, it is expressed as 
\begin{equation}
T=\begin{pmatrix}u^\ast_\uparrow&u^\ast_\downarrow&v^\ast_\downarrow&v^\ast_\uparrow\\-v_\uparrow&v_\downarrow&u_\downarrow&-u_\uparrow\end{pmatrix},
\end{equation}
with $(u_\uparrow,u_\downarrow,v_\downarrow,v_\uparrow)^T$ being the Nambu
spinors of $\psi^n$ at the point of the nanostructure connecting with lead 3.
Then, substituting Eqs.~(\ref{G_rApp}) and (\ref{GammaApp}) into Eq.~(\ref{Trans}),
the transmission coefficient for the case with the magnetic field along the ${\bf z}$-axis
is given by Eq.~(\ref{PehZ}) and that along the ${\bf x}$-axis is 
\begin{equation}
P^{eh}={\Gamma^e_L\Gamma^h_L}\left|{16(u_{\downarrow}-u_{\uparrow})(v^\ast_{\downarrow}+v^\ast_{\uparrow})\varepsilon}/{D(\varepsilon)}\right|^2,
\end{equation}
with $D(\varepsilon)=|u_{\downarrow}-u_{\uparrow}|^4\Gamma^e_L\Gamma^h_L-2|u_{\downarrow}-u_{\uparrow}|^2\{\Gamma^e_L\Gamma^h_L|v_{\downarrow}+v_{\uparrow}|^2+2i[\Gamma^h_L(\varepsilon-\varepsilon_n)+\Gamma^e_L(\varepsilon+\varepsilon_n)]\}-[4(\varepsilon+\varepsilon_n)+i\Gamma^e_L|v_{\downarrow}+v_{\uparrow}|^2][4(\varepsilon-\varepsilon_n)+i\Gamma^h_L|v_{\downarrow}+v_{\uparrow}|^2]$.

\begin{acknowledgments}
This work was supported by the National Natural Science Foundation of 
 China under Grant No. 11334014, the National Basic Research 
Program of China under Grant No. 2012CB922002 and the 
Strategic Priority Research Program of the Chinese
  Academy of Sciences under Grant No. XDB01000000. One of the authors (BYS) 
thanks Y. Zhou for valuable discussions. 

\end{acknowledgments}


\begin{thebibliography}{0}
\bibitem{Morestates} The differential conductances for the cases with
 $\phi=0$ and $0.1\phi_0$ shown in Fig.~\ref{figsw3}(b) do not exhibit peak
structure in the investigated energy range. This comes from 
the contribution of the
    third lowest state, whose energy is also small when the ring radius
 is large  ($\varepsilon_3=0.046t$ for $\phi=0$ and $0.044t$ for $\phi=0.1\phi_0$ in this case).
\end{thebibliography}
\end{document}